\documentclass[manuscript, nonacm]{acmart}
\AtBeginDocument{%
  }

\copyrightyear{2025}
\acmYear{2025}

\acmJournal{JACM}
\acmVolume{37}
\acmNumber{4}
\acmArticle{111}
\acmMonth{8}

\usepackage{hyperref}

\begin{document}

\title{Factually: Exploring Wearable Fact-Checking for Augmented Truth Discernment}

\begingroup
\renewcommand\thefootnote{}\footnote{
 This paper was presented at the 2025 ACM Workshop on Human-AI Interaction for Augmented Reasoning (AIREASONING-2025-01). This is the authors’ version for arXiv.}
\endgroup


\author{Chitralekha Gupta}
\email{chitralekha@nus.edu.sg}
\author{Hanjun Wu}
\email{michelle@ahlab.org}

 \author{Praveen Sasikumar}
 \email{praveen@ahlab.org}

 \author{Shreyas Sridhar}
\email{shreyas@ahlab.org}

 \author{Priambudi  Bagaskara}
\email{bagas@ahlab.org}

\author{Suranga Nanayakkara}
\email{suranga@ahlab.org}
\affiliation{%
  \institution{Augmented Human Lab, National University of Singapore}
  \country{Singapore}
}










  \begin{teaserfigure}
  \vspace{10pt}
    \includegraphics[width=0.39\textwidth]{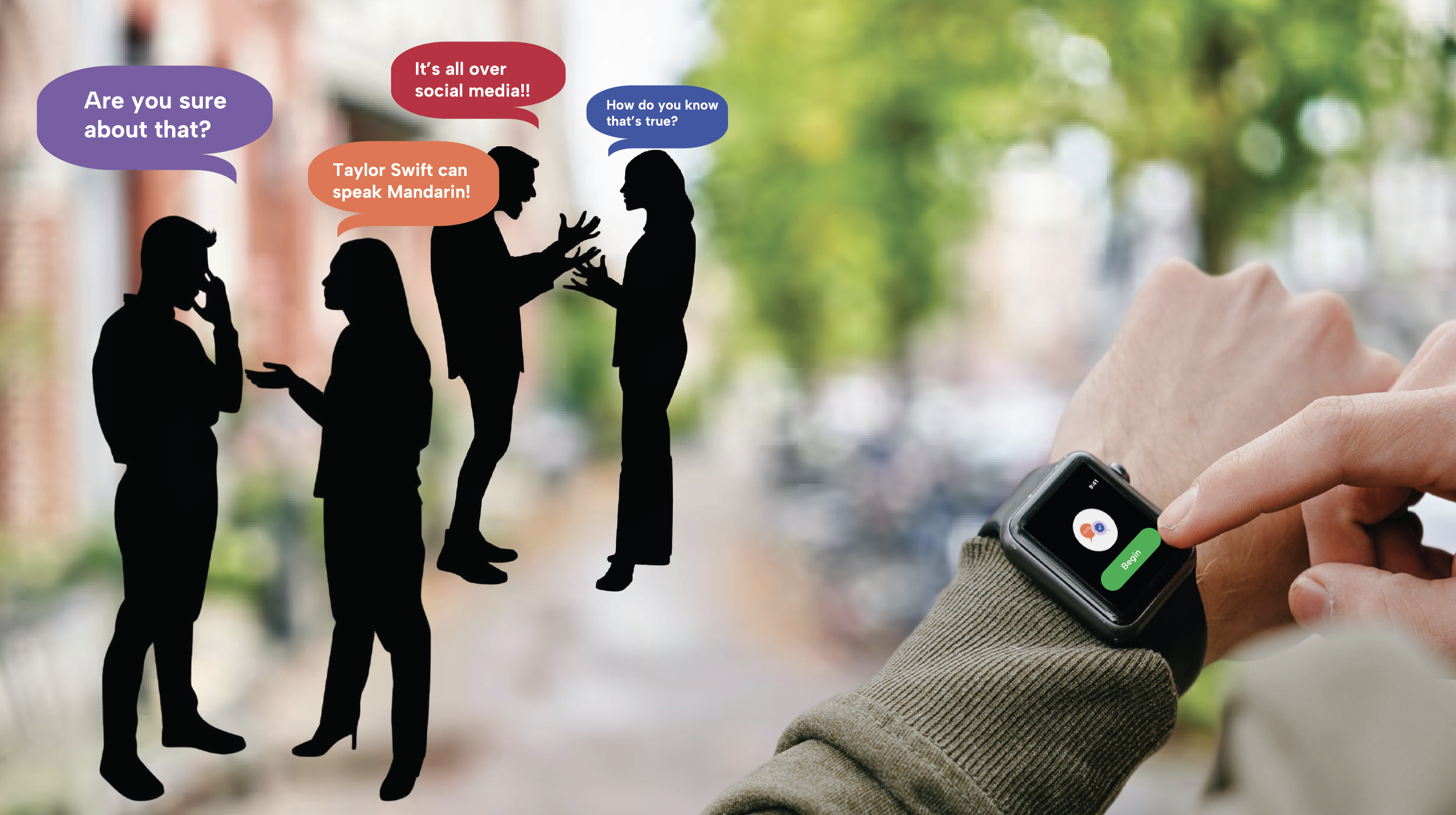}
    \includegraphics[width=0.3\textwidth]{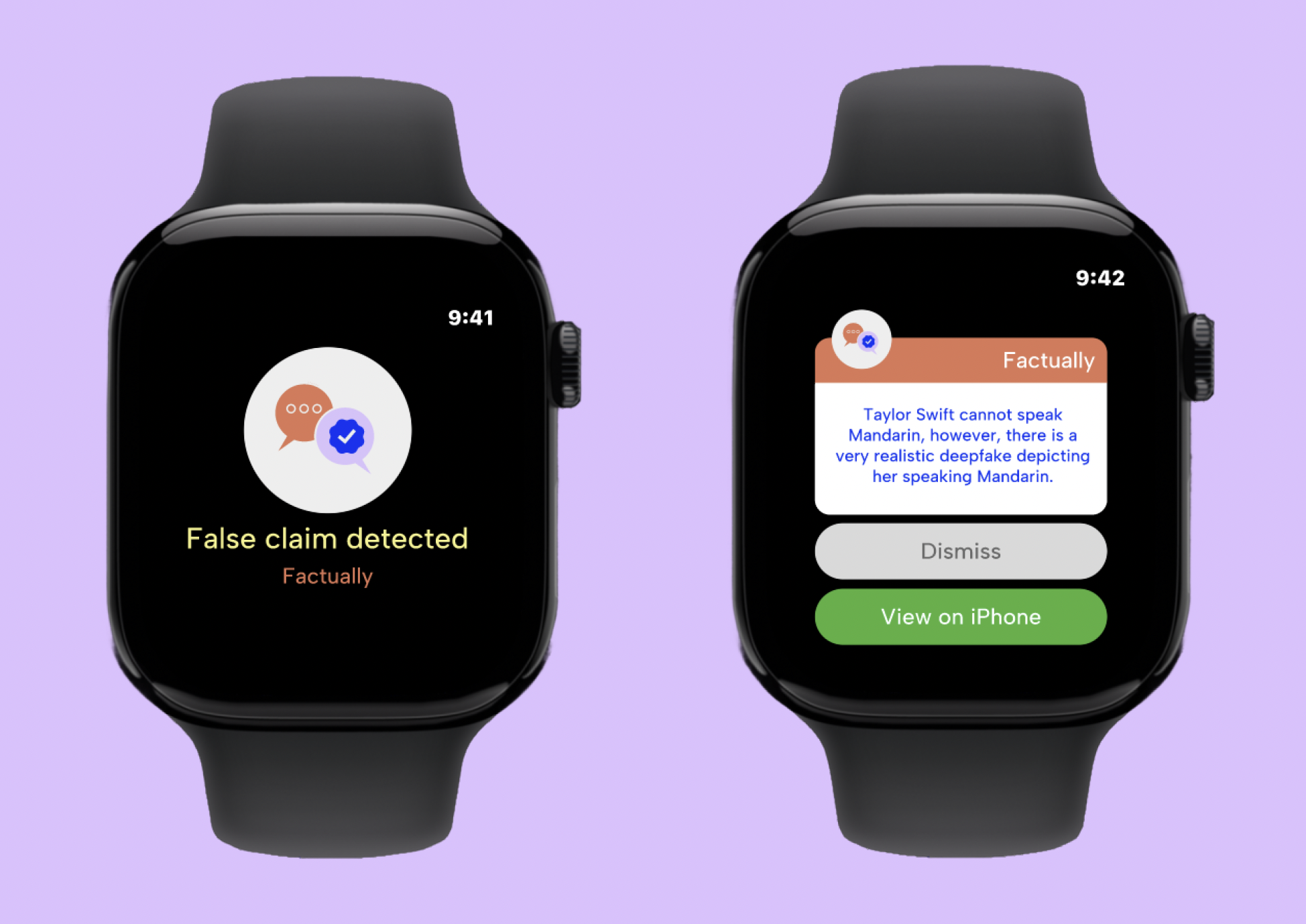}
    \includegraphics[width=0.3\textwidth]{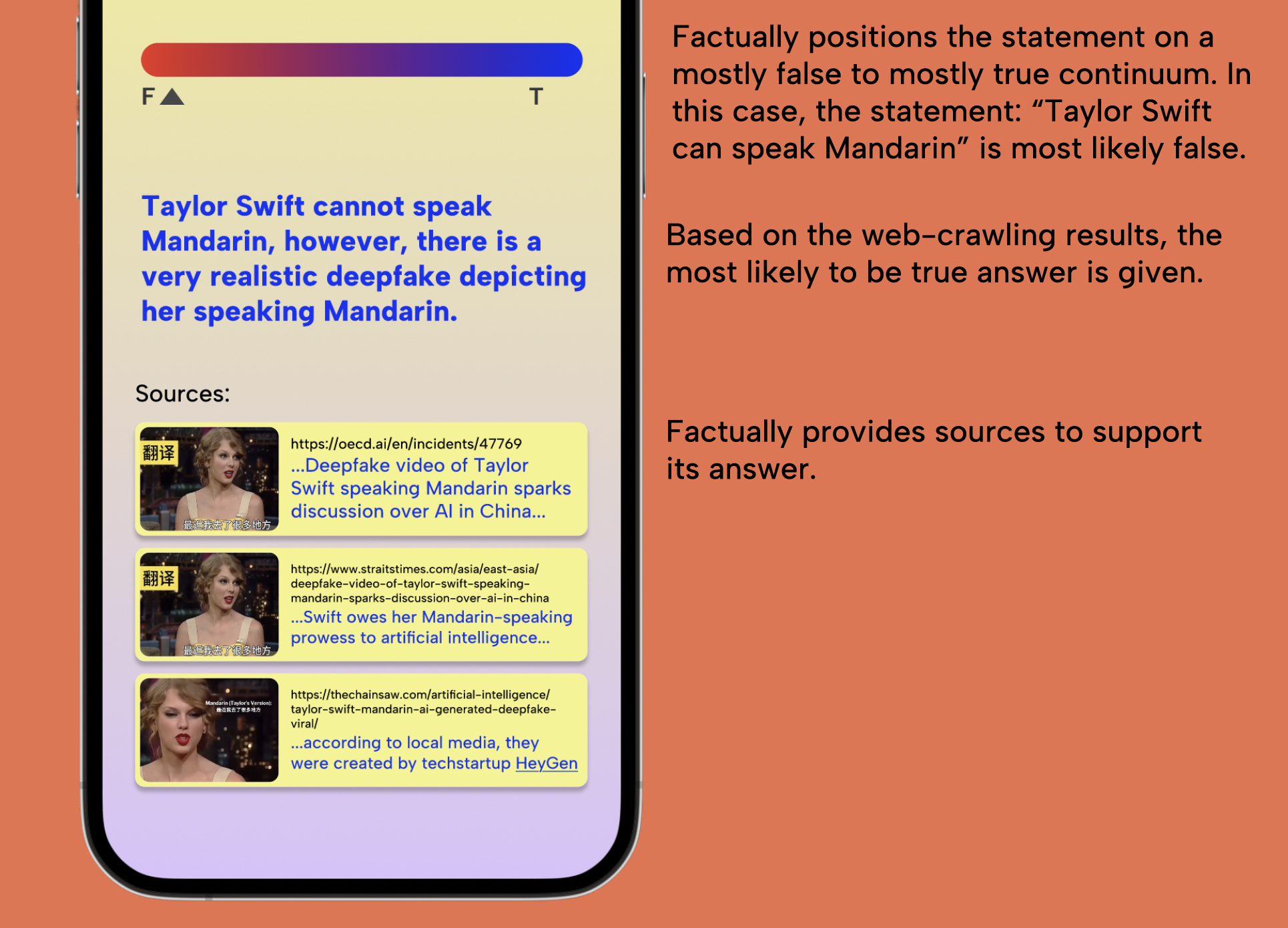}
    \caption{(a) A person using Factually in an everyday setting (b) Factually integrated with a smartwatch, (c) A companion mobile app.}
    \Description{(a) A person wearing a smart watch with the factually system downloaded. He is about to start fact checking. (b) Factually integrated with a smartwatch, (c) A companion mobile app.}
    \vspace{10pt}
    \label{fig:teaser}
  \end{teaserfigure}

\begin{abstract}
Wearable devices are transforming human capabilities by seamlessly augmenting cognitive functions. In this position paper, we propose a voice-based, interactive learning companion designed to amplify and extend cognitive abilities through informal learning. Our vision is threefold: (1) to enable users to discover new knowledge on-the-go through contextual interactive quizzes, fostering critical thinking and mindfulness; (2) to proactively detect misinformation, empowering users to critically assess information in real time; and (3) to provide spoken language correction and prompting hints for second language learning and effective communication. As an initial step toward this vision, we present Factually — a proactive, wearable fact-checking system integrated into devices like smartwatches or rings. Factually discreetly alerts users to potential falsehoods via vibrotactile feedback, helping them assess information critically. We demonstrate its utility through three illustrative scenarios\footnote{\href{https://drive.google.com/file/d/156BBsLr3YmJ8VOG-WVd6yhd5lCCTftt1/view?usp=sharing}{Concept Video Link}}, highlighting its potential to extend cognitive abilities for real-time misinformation detection. Early qualitative feedback suggests that Factually can enhance users' fact-checking capabilities, offering both practical and experiential benefits. 
\end{abstract}


\ccsdesc[500]{Human-centered computing~Haptic devices}
\ccsdesc[500]{Human-centered computing~Mobile devices}
\keywords{Assistive Augmentation, Fact-Checking, Wearable Assistant}


\maketitle

\section{Introduction}
Misinformation has become an unavoidable challenge in the digital age, with bite-sized falsehoods spreading rapidly across social media platforms and casual conversations. Fact-checking in real-time is inherently inconvenient, often socially awkward, and prone to being forgotten. Despite the rise of automated fact-checking systems, existing solutions typically require active engagement, such as searching for questionable claims, which disrupts natural conversations. To address this gap, we present \textit{Factually}, a wearable live fact-checking system designed to augment an individual’s ability to discern truth seamlessly.

Factually integrates with everyday wearable devices such as smartwatches or rings to provide real-time, discreet feedback through vibrations when potentially false information is detected. The aim of  \textit{Factually} aligns with the goals of assistive augmentation \cite{tan2025assistive} where it enhances cognitive capabilities in naturalistic settings, blending immediacy and social integration into an intuitive system. Our work builds on the broader vision of using wearable technologies to amplify human abilities, with a specific focus on tackling misinformation.

In this paper, we describe the design of \textit{Factually}, its implementation using existing fact-checking mechanisms, and its potential to transform how individuals engage with information. Through initial qualitative demonstrations, we highlight \textit{Factually}'s promise as an effective tool for fostering truth-centered behaviors and enhancing human cognition in socially integrated ways.

\section{Related Work}
\textit{Factually} builds upon a growing body of research in misinformation detection, assistive augmentation \cite{tan2025assistive}, and wearable technologies, extending these fields by addressing the challenges of real-time fact-checking in social contexts.

Several key studies have advanced our understanding of misinformation, particularly the cognitive and social processes that underpin its spread and persistence. 
Lewandowsky et al.~have made significant contributions to this field, exploring why individuals believe misinformation and the challenges of correcting it. 
For instance, \cite{ecker2010explicit} demonstrated that while explicit warnings can reduce the impact of misinformation, they often fail to eliminate its lasting effects. 
Similarly, 
Blank's work on "double misinformation" \cite{blank2022double} shows how misinformation can confuse people and distort their understanding of events. Blank’s earlier work \cite {blank2016past} on the social construction of memory further explains how misinformation impacts the formation of false memories.
These impactful contributions, alongside studies like \cite{johnson1994sources} work on the "continued influence effect" and \cite{del2016spreading} analysis of misinformation dynamics online, emphasize the complex interaction of psychological and social factors in the spread of misinformation and provide valuable insights into strategies for mitigation.

\subsection{Misinformation Detection and Cognitive Augmentation}
Kozyreva et al.~\cite{kozyreva2020citizens,kozyreva2024toolbox} discuss interventions aimed at combating misinformation and emphasize the importance of timely cognitive support. Pennycook et al.~\cite{pennycook2021psychology} examine the role of cognitive reflection in discerning truth and demonstrate that nudges to encourage reflective thinking can mitigate the spread of misinformation. Through factually, we align with these principles by providing real-time fact-checking assistance, and employ subtle, tactile cues that prompt introspection and self-correction without disrupting conversations.

\subsection{Wearables for Misinformation Detection}
Wearable technologies have increasingly been employed as tools for augmenting human cognitive and sensory abilities. 
The Wearable Reasoner explored the use of explainable AI to enhance rational decision-making by providing verbal feedback through a wearable device \cite{danry2020wearable}. While effective in delivering justifications for decisions, its focus on audio-based interactions raises practical challenges in socially sensitive scenarios. Factually addresses these limitations by relying on non-verbal tactile feedback, ensuring seamless integration into conversations. Factually builds on the foundation of \textit{assistive augmentation} ~\cite{tan2025assistive}, which envisions wearable technologies as transformative tools for expanding human capabilities. 

\subsection{Real-time Fact Checking}
Fact-checking systems have traditionally focused on web-based or multimedia platforms, with limited attention to conversational contexts. For example, Rashkin et al. ~\cite{rashkin2017truth} propose a language model-based system to detect false claims in online text. While effective in structured environments, such solutions are less applicable to dynamic, interpersonal interactions. Factually bridges this gap by operating as a wearable assistant capable of on-the-fly fact verification during conversations. Similarly, Setty et al.~ \cite{setty2024livefc} introduce Factiverse, a live fact-checking tool for online applications. While Factiverse targets media platforms, Factually focuses on augmenting individual users' cognitive abilities in real-world, conversational scenarios. 



\section{\textit{Factually}}
\subsection{Design considerations in the Assistive Augmentation space}
The design of the  \textit{Factually} is inspoired by the principles of assistive augmentation, as outlined by Tan et al.~\cite{tan2025assistive}. They introduce assistive augmentation as a paradigm that enhances human capabilities along two primary dimensions: \textit{ability} (expanding cognitive and sensory capacities) and \textit{integration} (seamlessly embedding augmentation into everyday life). We design \textit{Factually} so that it embodies these dimensions in the context of misinformation detection and truth discernment, in the following way:

\begin{figure}
   \vspace{-10pt}
    \centering
    \includegraphics[width=0.8\linewidth]{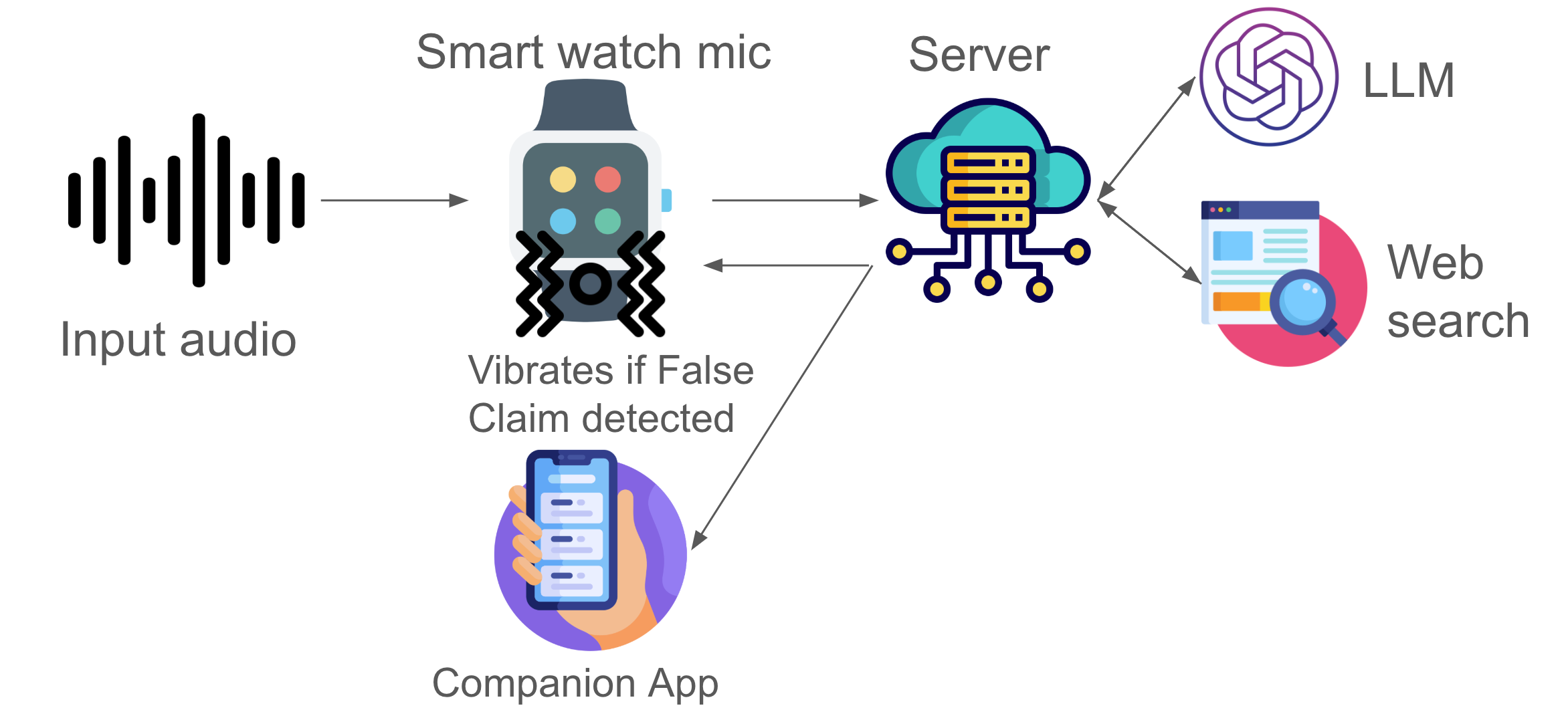}
    \caption{Overview of the Technical Implementation.} 
    \label{fig:system}
\end{figure}

\textit{Ability Dimension}: \textit{Factually} augments human cognitive capabilities by providing real-time feedback about potentially false information. It empowers users to discern truth more effectively during conversations, addressing the cognitive overload often associated with manual fact-checking. The system’s vibrational feedback mechanism enables users to extend their perceptual limits, allowing for quick and intuitive recognition of misinformation. This aligns with the ability dimension of assistive augmentation, where tools amplify human cognition without replacing human judgment.

\textit{Integration Dimension}:
The integration dimension emphasizes the importance of embedding augmentation seamlessly into users’ daily lives. \textit{Factually} achieves this through its wearable design, which uses discreet tactile feedback instead of disruptive auditory or visual cues. Whether integrated into a smartwatch or a smart ring, Factually ensures that users remain engaged in social interactions without drawing attention to the system’s operation. Its subtle design makes it particularly suitable for naturalistic settings, such as casual conversations or collaborative environments, where overt fact-checking would be impractical or socially awkward.


\subsection{System Design}
Our proof-of-concept system consists of the following components, as shown in Figure \ref{fig:system}:
\begin{itemize}
    \item Wearable Interface: Factually integrated into common wearable device (e.g. smartwatch). Vibrotactile feedback is used to indicate when a statement is flagged as potentially false, ensuring subtle and socially acceptable interactions.
    \item Fact-Checking Backend: The system leverages a combination of large language models (LLMs) and web-based resources to evaluate the truthfulness of statements in real time. While the current implementation uses state-of-the-art fact-checking tools as a proof of concept, this mechanism can be replaced with more sophisticated models as they become available.
    \item Real-Time Processing: Audio input is transcribed and analyzed for potentially false claims. If flagged, Factually sends a vibration alert to the wearable device. Users can then access additional context about the flagged statement through a connected mobile application, if desired.
\end{itemize}

\subsection{Proof-of-concept Scenarios}
\begin{figure}
   \vspace{-10pt}
    \centering    \includegraphics[width=\linewidth]{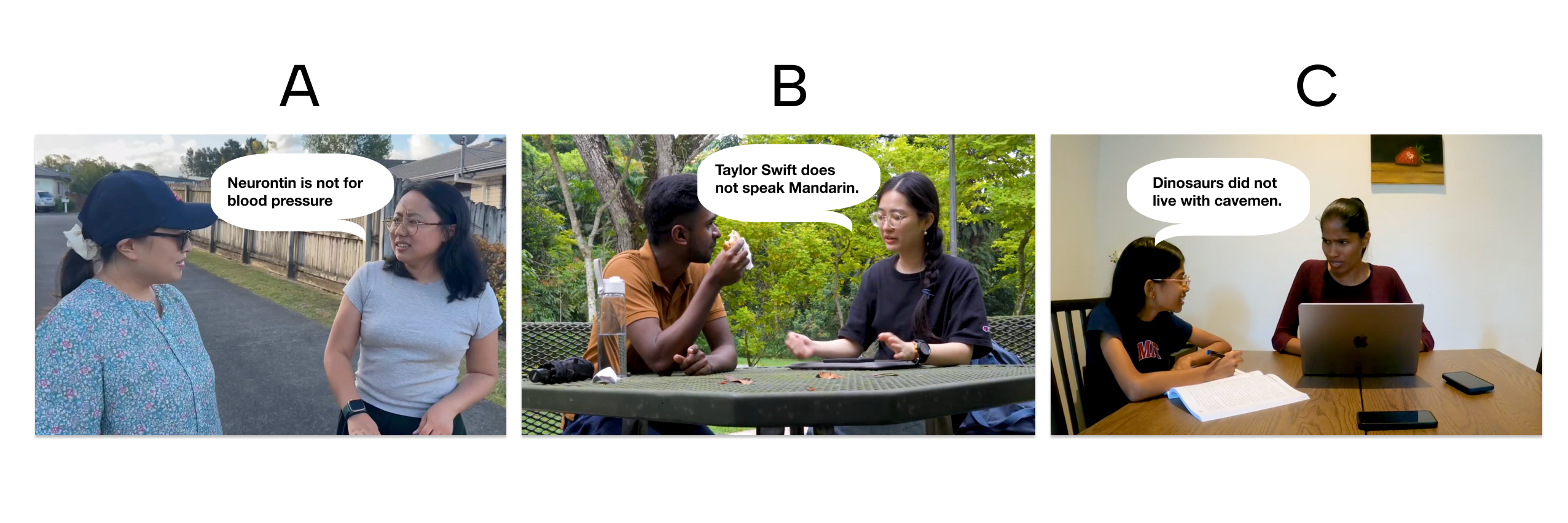}
   \vspace{-10pt}
    \caption{Use-Case Scenarios of Factually. (A) Scenario 1: health-related misinformation detection during a conversation between two grand daughters. (B) Scenario 2: misinformation detection during a casual lunch conversation. (C)  Scenario 3: misinformation detection during learning.}
    \label{fig:scenarios}
    \vspace{-10pt}
\end{figure}

To demonstrate the potential of \textit{Factually}, we developed three proof-of-concept scenarios (Figure \ref{fig:scenarios})\footnote{\href{https://drive.google.com/file/d/156BBsLr3YmJ8VOG-WVd6yhd5lCCTftt1/view?usp=sharing}{Concept Video Link}}.

\subsubsection{Scenario 1: Health Misinformation} 
In this scenario, we assess the role of Factually in detecting health-related misinformation in a live conversation. This scenario has two granddaughters, Emma and Grace, who are taking out the rubbish while discussing their grandmother’s medication.

Emma asks Grace if she has given their grandmother Neurontin. Grace responds that she hasn’t and questions the purpose of the medication. Emma confidently states that it is used to lower blood pressure.

At this point, \textit{Factually} vibrates, alerting the users to a potential error. The device provides a tactile cue, prompting the wearer to check the accompanying feedback: Neurontin is a medication used to help manage epileptic seizures. With this information, Grace points out the error, avoiding a potential health risk.

This scenario demonstrates \textit{Factually}’s ability to extend perceptual capabilities by identifying incorrect statements and providing accurate information in real time. By doing so, it enhances user awareness and helps prevent potentially harmful misunderstandings.

\subsubsection{Scenario 2: Social Conversations}
In this scenario, we assess the role of factually in a casual social setting. Two friends are having lunch and discussing Taylor Swift. One claims that Taylor Swift can speak Chinese, while the other disagrees, leading to a playful bet: the loser will pay for lunch.

The first individual finds a YouTube video appearing to show Taylor Swift speaking Mandarin. However, \textit{Factually} vibrates subtly, allowing the user to access the feedback: while Taylor Swift does not speak Mandarin, there is a highly realistic deepfake depicting her doing so. Armed with this information, the user refutes the claim, shifting the outcome of the bet.

This scenario illustrates how \textit{Factually}’s wearable and tactile design supports its seamless integration into social interactions. By aligning with familiar gestures, such as glancing at a wrist or adjusting a wearable device, \textit{Factually} remains inconspicuous while providing timely and actionable information.

\subsubsection{Scenario 3: Everyday Learning}
In this scenario, we evaluate the role of \textit{Factually} in learning. A parent is helping their child with homework. The child asks if dinosaurs lived alongside humans. The parent confidently responds affirmatively.

\textit{Factually} vibrates, prompting the parent to consult the device’s feedback: dinosaurs went extinct 65 million years before humans appeared. Armed with this new information, the parent corrects their statement, encouraging a more accurate understanding of history.

This scenario highlights how \textit{Factually} encourages users to question the validity of their claims and practice mindfulness in speech. By promoting introspection and self-correction, Factually fosters a truth-centered identity and critical thinking.
\subsection{Initial User Reactions}
We presented \textit{Factually} to 10 potential users to get their initial qualitative feedback. 
%
%
%
%
%
Overall, participants found \textit{Factually} intuitive and useful, particularly appreciating its non-intrusive nature. They expressed interest in using such a system in their daily lives, citing its potential to enhance critical thinking and mindfulness. However, they also suggested improvements in the vibrotactile feedback's specificity and the ability to customize the system's sensitivity to different topics.

\section{Limitations and Future Work}
The current prototype of \textit{Factually}  employs general-purpose large language models and web-based resources, which may not always produce domain-specific or highly accurate results. Future implementations would incorporate specialized fact-checking models trained on misinformation datasets to enhance reliability. 

\textit{Factually}  relies on cloud-based fact-checking using large language models and web resources, but network latency can delay real-time feedback. Future versions could address this by incorporating edge computing or on-device inference for faster response times. 

While \textit{Factually} ’s discreet vibrotactile feedback minimizes social disruptions, broader and long-term usability studies are needed to assess its practicality in diverse social contexts to guage its social acceptibility as well as its impact on user's cognitive load. Investigating the ethical implications of real-time fact-checking systems, such as privacy concerns and the potential for misuse, will be vital to ensure responsible deployment.

\bibliographystyle{ACM-Reference-Format}
\bibliography{sample-acmsmall}


\begin{thebibliography}{12}


\ifx \showCODEN    \undefined \def \showCODEN     #1{\unskip}     \fi
\ifx \showDOI      \undefined \def \showDOI       #1{#1}\fi
\ifx \showISBNx    \undefined \def \showISBNx     #1{\unskip}     \fi
\ifx \showISBNxiii \undefined \def \showISBNxiii  #1{\unskip}     \fi
\ifx \showISSN     \undefined \def \showISSN      #1{\unskip}     \fi
\ifx \showLCCN     \undefined \def \showLCCN      #1{\unskip}     \fi
\ifx \shownote     \undefined \def \shownote      #1{#1}          \fi
\ifx \showarticletitle \undefined \def \showarticletitle #1{#1}   \fi
\ifx \showURL      \undefined \def \showURL       {\relax}        \fi
\providecommand\bibfield[2]{#2}
\providecommand\bibinfo[2]{#2}
\providecommand\natexlab[1]{#1}
\providecommand\showeprint[2][]{arXiv:#2}

\bibitem[Blank et~al\mbox{.}(2022)]%
        {blank2022double}
\bibfield{author}{\bibinfo{person}{Hartmut Blank}, \bibinfo{person}{Anu Panday}, \bibinfo{person}{Ross Edwards}, \bibinfo{person}{Ewa Skopicz-Radkiewicz}, \bibinfo{person}{Violet Gibson}, {and} \bibinfo{person}{Vasudevi Reddy}.} \bibinfo{year}{2022}\natexlab{}.
\newblock \showarticletitle{Double misinformation: Effects on eyewitness remembering.}
\newblock \bibinfo{journal}{\emph{Journal of Applied Research in Memory and Cognition}} \bibinfo{volume}{11}, \bibinfo{number}{1} (\bibinfo{year}{2022}), \bibinfo{pages}{97}.
\newblock


\bibitem[Blank et~al\mbox{.}(2016)]%
        {blank2016past}
\bibfield{author}{\bibinfo{person}{Hartmut Blank}, \bibinfo{person}{Eva Walther}, {and} \bibinfo{person}{Simon~D Isemann}.} \bibinfo{year}{2016}\natexlab{}.
\newblock \showarticletitle{The past is a social construction: susceptibility to social inf luence in (mis) remembering}.
\newblock In \bibinfo{booktitle}{\emph{False and distorted memories}}. \bibinfo{publisher}{Psychology Press}, \bibinfo{pages}{65--81}.
\newblock


\bibitem[Danry et~al\mbox{.}(2020)]%
        {danry2020wearable}
\bibfield{author}{\bibinfo{person}{Valdemar Danry}, \bibinfo{person}{Pat Pataranutaporn}, \bibinfo{person}{Yaoli Mao}, {and} \bibinfo{person}{Pattie Maes}.} \bibinfo{year}{2020}\natexlab{}.
\newblock \showarticletitle{Wearable Reasoner: towards enhanced human rationality through a wearable device with an explainable AI assistant}. In \bibinfo{booktitle}{\emph{Proceedings of the Augmented Humans International Conference}}. \bibinfo{pages}{1--12}.
\newblock


\bibitem[Del~Vicario et~al\mbox{.}(2016)]%
        {del2016spreading}
\bibfield{author}{\bibinfo{person}{Michela Del~Vicario}, \bibinfo{person}{Alessandro Bessi}, \bibinfo{person}{Fabiana Zollo}, \bibinfo{person}{Fabio Petroni}, \bibinfo{person}{Antonio Scala}, \bibinfo{person}{Guido Caldarelli}, \bibinfo{person}{H~Eugene Stanley}, {and} \bibinfo{person}{Walter Quattrociocchi}.} \bibinfo{year}{2016}\natexlab{}.
\newblock \showarticletitle{The spreading of misinformation online}.
\newblock \bibinfo{journal}{\emph{Proceedings of the national academy of Sciences}} \bibinfo{volume}{113}, \bibinfo{number}{3} (\bibinfo{year}{2016}), \bibinfo{pages}{554--559}.
\newblock


\bibitem[Ecker et~al\mbox{.}(2010)]%
        {ecker2010explicit}
\bibfield{author}{\bibinfo{person}{Ullrich~KH Ecker}, \bibinfo{person}{Stephan Lewandowsky}, {and} \bibinfo{person}{David~TW Tang}.} \bibinfo{year}{2010}\natexlab{}.
\newblock \showarticletitle{Explicit warnings reduce but do not eliminate the continued influence of misinformation}.
\newblock \bibinfo{journal}{\emph{Memory \& cognition}}  \bibinfo{volume}{38} (\bibinfo{year}{2010}), \bibinfo{pages}{1087--1100}.
\newblock


\bibitem[Johnson and Seifert(1994)]%
        {johnson1994sources}
\bibfield{author}{\bibinfo{person}{Hollyn~M Johnson} {and} \bibinfo{person}{Colleen~M Seifert}.} \bibinfo{year}{1994}\natexlab{}.
\newblock \showarticletitle{Sources of the continued influence effect: When misinformation in memory affects later inferences.}
\newblock \bibinfo{journal}{\emph{Journal of experimental psychology: Learning, memory, and cognition}} \bibinfo{volume}{20}, \bibinfo{number}{6} (\bibinfo{year}{1994}), \bibinfo{pages}{1420}.
\newblock


\bibitem[Kozyreva et~al\mbox{.}(2020)]%
        {kozyreva2020citizens}
\bibfield{author}{\bibinfo{person}{Anastasia Kozyreva}, \bibinfo{person}{Stephan Lewandowsky}, {and} \bibinfo{person}{Ralph Hertwig}.} \bibinfo{year}{2020}\natexlab{}.
\newblock \showarticletitle{Citizens versus the internet: Confronting digital challenges with cognitive tools}.
\newblock \bibinfo{journal}{\emph{Psychological Science in the Public Interest}} \bibinfo{volume}{21}, \bibinfo{number}{3} (\bibinfo{year}{2020}), \bibinfo{pages}{103--156}.
\newblock


\bibitem[Kozyreva et~al\mbox{.}(2024)]%
        {kozyreva2024toolbox}
\bibfield{author}{\bibinfo{person}{Anastasia Kozyreva}, \bibinfo{person}{Philipp Lorenz-Spreen}, \bibinfo{person}{Stefan~M Herzog}, \bibinfo{person}{Ullrich~KH Ecker}, \bibinfo{person}{Stephan Lewandowsky}, \bibinfo{person}{Ralph Hertwig}, \bibinfo{person}{Ayesha Ali}, \bibinfo{person}{Joe Bak-Coleman}, \bibinfo{person}{Sarit Barzilai}, \bibinfo{person}{Melisa Basol}, {et~al\mbox{.}}} \bibinfo{year}{2024}\natexlab{}.
\newblock \showarticletitle{Toolbox of individual-level interventions against online misinformation}.
\newblock \bibinfo{journal}{\emph{Nature Human Behaviour}} (\bibinfo{year}{2024}), \bibinfo{pages}{1--9}.
\newblock


\bibitem[Pennycook and Rand(2021)]%
        {pennycook2021psychology}
\bibfield{author}{\bibinfo{person}{Gordon Pennycook} {and} \bibinfo{person}{David~G Rand}.} \bibinfo{year}{2021}\natexlab{}.
\newblock \showarticletitle{The psychology of fake news}.
\newblock \bibinfo{journal}{\emph{Trends in cognitive sciences}} \bibinfo{volume}{25}, \bibinfo{number}{5} (\bibinfo{year}{2021}), \bibinfo{pages}{388--402}.
\newblock


\bibitem[Rashkin et~al\mbox{.}(2017)]%
        {rashkin2017truth}
\bibfield{author}{\bibinfo{person}{Hannah Rashkin}, \bibinfo{person}{Eunsol Choi}, \bibinfo{person}{Jin~Yea Jang}, \bibinfo{person}{Svitlana Volkova}, {and} \bibinfo{person}{Yejin Choi}.} \bibinfo{year}{2017}\natexlab{}.
\newblock \showarticletitle{Truth of varying shades: Analyzing language in fake news and political fact-checking}. In \bibinfo{booktitle}{\emph{Proceedings of the 2017 conference on empirical methods in natural language processing}}. \bibinfo{pages}{2931--2937}.
\newblock


\bibitem[Setty et~al\mbox{.}(2024)]%
        {setty2024livefc}
\bibfield{author}{\bibinfo{person}{Vinay Setty} {et~al\mbox{.}}} \bibinfo{year}{2024}\natexlab{}.
\newblock \showarticletitle{LiveFC: A System for Live Fact-Checking of Audio Streams}.
\newblock \bibinfo{journal}{\emph{arXiv preprint arXiv:2408.07448}} (\bibinfo{year}{2024}).
\newblock


\bibitem[Tan et~al\mbox{.}(2025)]%
        {tan2025assistive}
\bibfield{author}{\bibinfo{person}{Felicia Fang-Yi Tan}, \bibinfo{person}{Chitralekha Gupta}, \bibinfo{person}{Dixon Prem~Daniel Rajendran}, \bibinfo{person}{Pattie Maes}, {and} \bibinfo{person}{Suranga Nanayakkara}.} \bibinfo{year}{2025}\natexlab{}.
\newblock \showarticletitle{Assistive Augmentation: Fundamentally Transforming Human Ability}.
\newblock \bibinfo{journal}{\emph{Interactions}} \bibinfo{volume}{32}, \bibinfo{number}{1} (\bibinfo{year}{2025}), \bibinfo{pages}{22--27}.
\newblock


\end{thebibliography}

\appendix

\end{document}